\documentclass[aps,twocolumn,groupedaddress,showpacs,showkeys]{revtex4}%
\usepackage{amsmath}
\usepackage{graphicx}%
\usepackage{amsfonts}%
\usepackage{amssymb}

\begin{document}
\bibliographystyle{apsrev}

\title{Black chain of pearls in 5D de Sitter spacetime
\footnote{Work supported by 
the National Natural Science Foundation of
China through grant No. 90403014.}
}
\author{Liu Zhao}
\email{lzhao@nankai.edu.cn}
\affiliation{Department of Physics, Nankai University, 
Tianjin 300071, P R China}

\date{\today}

\begin{abstract}
We analyze some exact chain-shaped black hole solutions in 5-dimensional spacetime. Unlike usual 
black string and black ring solutions, the topology of the horizons of the new solutions 
are neither $\mathbb{R}\times R^{2}$ nor $S^1\times S^2$ but rather like several topological spheres 
concatenating each other at single points. The shape of the horizon suggests the name \emph{black chain 
of pearls} on which each \emph{pearl} is a topological $3$-sphere on the chain.  
In addition to the usual black hole hairs, the number of pearls can be viewed as a new hair of 
the black chain of pearls.
\end{abstract}
\pacs{04.50.+h, 04.20.Gz, 04.70.-s}
\keywords{black hole, horizon topology }

\maketitle

\vspace{0.5cm} 

Black holes are a fascinating class of solutions appearing in Einstein theory of gravity 
as well as many modified theories of gravity. In 4 spacetime dimensions there exist a number of 
uniqueness theorems which guarantee the uniqueness of black hole solution under a given set of 
conditions. In higher dimensions black hole uniqueness does not hold and a number of exotic 
black hole solutions, known as black rings, were found \cite{hep-th/0110258, hep-th/0110260,
hep-th/0412153} in the last few years. These black rings 
have horizons of $S^1 \times S^2$ topology and are mostly living in 5 dimensional asymptotically 
flat spacetime, usually carrying spin \cite{hep-th/0110260} and/or 
electro-magnetic charges \cite{hep-th/0412153} in order to avoid 
conical singularities. One astonishing property of the black ring solutions is that they
can be more stable than the corresponding spherical black holes with the same mass, 
charge and spin. Most recently it was found \cite{EF, EEF} that a particular black hole 
configuration called black Saturn which consists of a spherical black hole surrounded 
by a thin black ring is even more stable than black rings. The existence of all 
these unusual black solutions in 5 dimensions 
makes the phase diagram of 5D gravity an active field of study. 

In a recent paper by Chu and Dai \cite{hep-th/0611325}, a 5-dimensional (5D) ``black
ring'' solution with positive cosmological constant was constructed. The
construction is based on the warped decomposition of the 5D
geometry in terms of the $4$-dimensional charged de Sitter C-metric, i.e.%
\begin{align}
ds_{5}^{2}  &  =dz^{2}+\cos^{2}(kz)ds_{4}^{2},\label{1}\\
ds_{4}^{2}  &  =\frac{1}{A^{2}(x-y)^{2}}\left[  G(y)dt^{2}-\frac{dy^{2}}%
{G(y)}+\frac{dx^{2}}{\tilde{G}(x)}+\tilde{G}(x)d\varphi^{2}\right]  ,\nonumber
\end{align}
where%
\begin{align*}
G(\xi)  &  =q^{2}A^{2}\xi^{4}+a_{3}\xi^{3}+a_{3}\xi^{2}+a_{1}\xi+a_{0},\\
\tilde{G}(\xi)  &  =G(\xi)-k^{2}/A^{2},
\end{align*}
and $ds_{4}^{2}$ must be accompanied by a Maxwell potential%
\[
A_{\varphi}=qx+c_{0}%
\]
in order to constitute a solution to the Einstein-Maxwell equation in the
presence of a cosmological constant. The argument which makes the solution
(\ref{1}) a black ring is as follows. First, the metric $ds_{4}^{2}$, as a de
Sitter generalization of the well known C-metric, corresponds to two black
holes accelerating apart in 4 dimensions. Then, since the warped factor
$\cos^{2}(kz)$ is periodic in $z$, the horizon topology becomes $S^{1}\times
S^{2}$ in 5 dimensions. So the standard method for analyzing black rings in
asymptotically flat spacetime can be systematically applied to the de Sitter
black ring which constitutes the major part of \cite{hep-th/0611325}.

In this article we first point out that actually the horizon topology of warped
geometries like (\ref{1}) is not $S^{1}\times S^{2}$, because at $kz=\frac
{\pi}{2}\operatorname{mod}n\pi$, the $S^{2}$ factor in the horizon shrinks
to zero size and at those points the local geometry of the horizons look like two 
cones concatenating each other at the tops. 
Moreover, since the metric and the horizon is not translationally invariant
along the $z$ axis, it is not necessary to require the coordinate $z$ to
extend only over a single period: allowing $z$ to extend over several periods
of $\cos(kz)$ would correspond to very different horizon topologies. Despite
the above remarks, we would like to adopt the construction used in
\cite{hep-th/0611325} to illustrate some simpler axial symmetric de Sitter
solutions in 5-dimensions.

Our constructions will be based on the following formula which also appeared
in \cite{hep-th/0611325}: a generic $D$-dimensional metric with positive
cosmological constant can be embedded into a $(D+1)$-dimensional geometry also
with positive cosmological constant via%
\begin{equation}
ds_{(D+1)}^{2}=dz^{2}+\cos^{2}(kz)ds_{D}^{2}, \label{2}
\end{equation}
and the corresponding Ricci tensors are related by%
\begin{align*}
\hat{R}_{\mu \nu}-Dk^{2}\hat{g}_{\mu \nu} &  =R_{\mu \nu}-(D-1)k^{2}g_{\mu \nu
},\\
\hat{R}_{zz} &  =Dk^{2}.
\end{align*}
Now at $D=4$, instead of de Sitter C-metric, we insert the usual 4-dimensional
Schwarzschild-de Sitter metric%
\begin{align}
ds_{4}^{2} &  =-f(r)dt^{2}+\frac{1}{f(r)}dr^{2}+r^{2}(d\theta^{2}+\sin
^{2}\theta d\varphi^{2}), \label{sds}\\
f(r) &  =1-\frac{2M}{r}-\frac{\Lambda r^{2}}{3},\qquad \Lambda=3k^{2} \label{f}
\end{align}
into (\ref{2}). The resulting 5D metric $ds_{5}^{2}$ will then be a
solution to the vacuum Einstein equation with positive cosmological constant
$4k^{2}$. As mentioned earlier, since the metric is periodic but not
translationally invariant along $z$, we have no reason to restrict $z$ to take
values in only a single period. Instead, we allow $z$ to take values in the
range of several integral multiples of the period of the metric functions.
This will not affect the local geometry but generically will change the
topology of the horizon. To make it more explicit, one can caculate the curvature invariants of the metric 
described by (\ref{2}), (\ref{sds}) and (\ref{f}). One of these turns out to be
\[
R^{\mu\nu\rho\sigma}R_{\mu\nu\rho\sigma} =8 \left( 5k^{4} + \frac{6M^{2}}{r^{6}\cos^{4}(k z)}\right).
\]
The singularity at $r=0$ is surrounded by roots $r_{\pm}$ of $f(r)$, i.e. the horizons, while the 
singularities at $\cos(k z)=0$ are naked. The metric at the horizons can be written as 
\[
ds_{H_\pm}^2 = dz^2 + \cos^2(kz) r_{\pm}^2\left(d\theta^2+\sin^2 \theta d\varphi^2\right).
\]
To illustrate the horizon topology,
we depict the outer horizon of the metric corresponding to
allowing $z$ to run over $4$ periods. Of course the number of periods $4$ is picked for random here, any 
integer number of periods as allowed. This horizon is $3$-dimensional 
and while depicting it we have omitted the angular coordinate $\varphi$ appearing in the 
4-dimensional Schwarzschild-de Sitter part. Such solutions will be referred to 
as \emph{black chain of pearls} due to the particular shape of the horizons. The $2$-dimensional
slices of the horizon at constant $z$ are almost all $2$-spheres corresponding to the Schwarzschild-de Sitter 
horizons except at the naked singularities located at $k z=n + \pi/2$. These naked singularities 
are referred to as \emph{nodes} on the chain, and the segments of the chain between 
two nodes are called a {\it pearl}. Note that the topology of each pearl is a $3$-sphere and the black chain 
horizon as a whole is topologically equivalent to several $3$-spheres concatenating at isolated nodes.
We postulate that the appearance 
of these nodes signifies that such solutions might correspond to the 
``critical stage'' for the gravitational phase transition from an array of 5-dimensional black 
holes or a uniform 5D black string into a final stable configuration, presumably something similar to a 
5D black ring, because there have been numerous 
arguments stating that the large black ring phase is thermodynamically more favorable than an array 
of black holes, and the uniform black string is unstable due to Gregory-Laflamme 
instability \cite{GL}. 
\begin{figure}[htb]
\begin{center}
\includegraphics[width=2.4in]{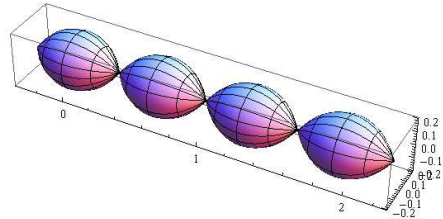}
\caption{Horizon of the black chain with 4 pearls} \label{Figure1}
\end{center}
\end{figure}

Besides different horizon topologies, there is another fundamental difference between the 
black chain of pearls and black strings/rings. As shown in Figure \ref{Figure1}, the black chain 
spacetimes are not translationally invariant in the fifth dimension (i.e. the $z$-direction). 
Instead, there is only a 
discrete symmetry in that direction which is $Z_p$, with $p$ the number of pearls on the chain. 
Gravitational solutions with discrete symmetries are always of great interests in the literature.

It is tempting to study the properties of the multi-period solution in detail. First comes the area 
of the horizon per unit period in $z$. We have%
\begin{align*}
A &  =4\pi r_{+}^{2}\int_{0}^{\pi/k}\cos^{2}(kz)dz\\
&  =\frac{2\pi^{2}r_{+}^{2}}{k}=2\pi^{2}r_{+}^{2}\ell,
\end{align*}
where $r_+$ is the radius of the outer horizon of the 4-dimensional Schwarzschild-de Sitter spacetime, and 
\[
\ell=\frac{1}{k}=\sqrt{\frac{3}{\Lambda}}
\]
is just the de Sitter raduis. The Bekenstein-Hawking entropy of the outer horizon then reads%
\[
S=\frac{p A}{4}=\frac{p \pi^{2}r_{+}^{2}\ell}{2}.
\]

To determine the temperature of the horizons, however, we need to take care of
a subtlety which generally appear in any black hole solutions admitting more
than one horizon. It is believed that such spacetimes are not in thermal
equilibrium and one can define Hawking temperatures separately if the
separation between the horizons is large enough and the shell of spacetime in
between the horizons can be thought of as adiabatic.  Under such assumptions it is an easy
practice to derive the Hawking temperature for the outer horizon, which turn out to be equal to the
Hawking temperature on the outer horizon of 4D Schwarzschild-de Sitter black hole under the same
assumptions \cite{T1,T2,T3},
\[
T=\frac{1}{2\pi\sqrt{1-(27y)^{1/3}}}\left|\frac{M}{r_{+}^{2}}-\frac{r_{+}}{\ell^{2}}
\right|,
\]
where $y=M^{2}/\ell^{2}$.

The black chain of pearls described in this article is of the simplest form, i.e. they are static and neutral. 
We can check that they are unstable against the gravitational perturbations to the 
first order. This can be done following the standard perturbation process usually pursued in the test of 
Gregory-Laflamme instability of various black strings in 5-dimensions. To do this, we first change the
metric into another gauge in which it looks as follows:
\[
ds_{5}^{2}=e^{B(\zeta)} \left(ds_{4}^{2}+d\zeta^{2}\right).
\]
It is easy to see that the desired coordinate transformation is
\[
z\rightarrow z(\zeta)= \frac{2}{k}\tan^{-1}\left(\tanh\left(\frac{k \zeta}{2}\right)\right),
\]
and the function $B(\zeta)$ reads
\[
B(\zeta)=\log(\mathrm{sech}^{2}(\zeta)).
\]
Then we can repeat the procedure of perturbation as we did recently in \cite{Zhao} to get the 
following set of perturbation equations:
\begin{align}
\left[  \square^{(\gamma)}h_{\mu \nu}(x)+2R_{\mu \rho \nu \lambda}(\gamma
)h^{\rho \lambda}(x)\right]   &  =m^{2}h_{\mu \nu}(x),\label{ho}\\
\left[  -\partial_{\zeta}^{2}+\frac{3k^{2}}{4}\left(3-5~ \mbox{sech}^2(k\zeta)\right) \right]  \xi(\zeta)  
&  =m^{2}\xi(\zeta), 
\label{sch}%
\end{align}
where the 5D metric $ds_5^2$ is perturbed as
\begin{align}
g_{MN}  &  \rightarrow g_{MN}+\delta g_{MN},\nonumber \\
d\hat{s}^{2}  &  \rightarrow d\hat{s}^{\prime2}=e^{B(\zeta)} \nonumber\\
& \times \left(  \left(
\gamma_{\mu \nu}+e^{-B(\zeta)} h_{\mu \nu}(x,\zeta)\right)  dx^{\mu}dx^{\nu}%
+d\zeta^{2}\right)  , \label{pert}%
\end{align}
$\gamma_{\mu\nu}$ represent components of the metric $ds_4^2$, and $m$ is a constant coming 
from the separation of variables which represents the energy scale of the perturbation. 
In the above, the metric perturbation $h_{\mu\nu}(x,\zeta)$ is related to $h_{\mu\nu}(x)$ 
and $\xi(\zeta)$ via
\[
h_{\mu \nu}(x,\zeta)=e^{B(\zeta)/4} h_{\mu \nu}(x) \xi(\zeta),
\]
and in order to cast the perturbation equations into the simple form (\ref{sch}) the following transverse
traceless and Lorentzian gauge condition is adopted, 
\begin{align*}
h  &  \equiv \gamma^{\mu \nu}h_{\mu \nu}=0,\\
\nabla_{\mu}h^{\mu \nu}  &  =0.
\end{align*}

\begin{figure}[htb]
\begin{center}
\includegraphics[width=2.4in]{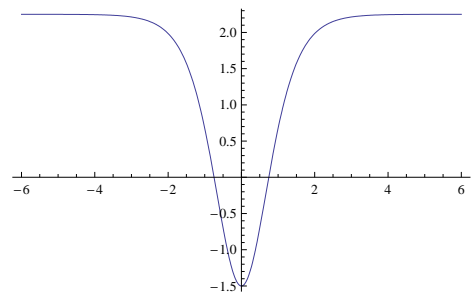}
\caption{Plot of the potential in (\ref{sch}) at $k=1$} \label{Figure3}
\end{center}
\end{figure}

The first of the perturbation equations (eq. (\ref{ho})) is the well known Lichnerowitz equation 
in 4-dimensional Schwarzschild-de Sitter background, which was shown to be always unstable by 
Hirayama and Kang in \cite{Hirayama-Kang, Kang}. The 
second of the perturbation equations (eq.(\ref{sch})) is a Schr\"odinger-like equation in the 
fifth dimension with a potential approaching a finite constant value $9k^{2}/4$ at large 
$\zeta$ (see Figure \ref{Figure3}). 
Thus the normalizable solutions thereof contain a spectrum with contineous eigenvalue $m^{2}$ and
hence any unstable modes with strength bigger than $9k^{2}/4$ will travel along $\zeta$ direction
without any barrier. Thus both parts of the perturbation equation signify instability of the solution.
In fact, such instability should already be forecasted when we first encountered the naked singularities 
at the nodes.

Notice, however, that the discussion made above using the $\zeta$ coordinate is a little bit flawed. This is 
because the $\zeta$ coordinate actually cannot represent the full range of $z$ values allowed in the solution.
In fact, $\zeta$ going from $-\infty$ to $\infty$ only corresponds to $kz$ going from $-\pi/2$ to $\pi/2$, 
i.e. only one period of $z$. Thus the ability for the instability modes to travel 
from $\zeta=-\infty$ to $\zeta=\infty$ does not mean that they
can travel all along the chain. Indeed, changing back to the $z$ coordinate, the second perturbation equation 
becomes
\[
\lbrack-\partial_{z}^{2}+2k^{2}\left(  \sec^{2}(kz)-2\right)  ]\eta
(z)=m^{2}\sec^{2}(kz)\eta(z),
\]
where $\eta(z)=\cos^{1/2}(kz)\xi(z)$. A similar equation also arised in \cite{Gregory}, in which 
stability of AdS black string is analysed. This equation does not look like a Schrodinger 
equation due to the apparence of the factor $\sec^{2}(kz)$ on the right hand side. 
Nevertheless, the potential term on the left
hand side is a confining function, thus the propogating modes along $z$ axis cannot travel through the nodes 
located at $kz=n+\pi/2$. In other words, the nodes cannot be blown up into $S^{2}$ with nonvanishing radius 
by the perturbation modes. Therefore the number $p$ of pearls on the chain (which equals the number of 
nodes plus one) is an invariant propery of the solution even in the presence of perturbation, so it might 
be regarded as yet another hair of the black chain solution. 

It will be interesting to make further understanding of the instability described above 
in the framework of Gregory-Laflamme instabilities. As pointed out by Horowitz and Maeda in 
\cite{Horowitz-Maeda}, black string (of which black ring is a particular branch) horizons 
cannot pinch off. So whatever the nature and strength the perturbation is,
a black string cannot segregate into black holes. Therefore, the black chain solution described 
in this article should not be broken into smaller segments after the perturbation. 

Combining the arguments made in the last two paragraphs, the black chains can neither be blown up 
into black strings without singular nodes, nor can they be broken into single pearls which are 
equivalent to short black strings with compact horizons. It remains to answer what is the final 
configuration of the black chains after the perturbation. 

It is tempting to find other black chain solutions with more complex properties, e.g. 
those carrying nontrivial electro-magnetic charges and/or rotation parameters, because in general
the inclusion of these parameters will improve the stability of the solution. To this end 
it should be mentioned that the solution found by Chu and Dai in \cite{hep-th/0611325} is 
actually a black di-chain carrying a pure magnetic charge if we allow $z$ to extend over several periods.

\begin{figure}[htb]
\begin{center}
\includegraphics[width=1.92in]{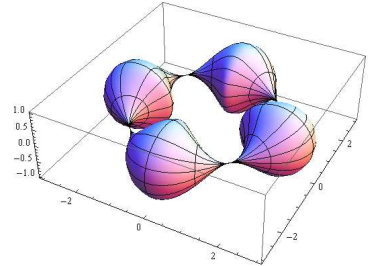}
\caption{Sketch of a black necklace} \label{Figure2}
\end{center}
\end{figure}

It is also tempting to find the closed analogues of the black chains (black ``necklaces'') 
which has horizons of the shape as skematically depicted in Figure \ref{Figure2}.
If ever such a configuration exist, it would arguably be more stable than the black chains described here, 
just as black rings are more stable than uniform black strings. However we need 
more direct evidence to justify the last statement.

Changing the 5D de Sitter bulk to spacetimes with different asymptotics will be another direction 
of further investigation. Last but not the least, it is also of great interests to compare the 
thermodynamic properties of black strings/rings and black chains/necklaces in detail. 
We hope to answer these questions in later works.

\vspace{0.3cm}


\end{document}